\documentclass[preprint]{ptephy_v1}

\preprintnumber{OU-HET-902, CTPU-16-20}

\usepackage{graphicx}
\usepackage{amsmath,amssymb}
\usepackage{bm}
\usepackage{color}
\usepackage{hyperref}
\hypersetup{ 
 setpagesize=false,
 bookmarksnumbered=true,
 colorlinks=true,
 linkcolor=blue,
 citecolor=red,
}

\newcommand{\eq}[1]{Eq.~(\ref{#1})}
\newcommand{\Sec}[1]{Sec.~\ref{#1}}
\newcommand{\ReC}[1]{(\text{Re} C_{#1})}
\newcommand{\AbsC}[1]{|C_{#1}|}

\begin{document}

\title{\boldmath New physics contributions in $B\to\pi\tau\bar\nu$ and $B\to\tau\bar\nu$}

\author[1]{Minoru TANAKA}
\author[2]{Ryoutaro WATANABE \thanks{These authors contributed equally to this work}}

\affil[1]{Department of Physics, Graduate School of Science, Osaka University,  Toyonaka, Osaka 560-0043, Japan \email{tanaka@phys.sci.osaka-u.ac.jp}} 
\affil[2]{Center for Theoretical Physics of the Universe, Institute for Basic Science (IBS), Daejeon, 34051, Republic of Korea \email{wryou1985@ibs.re.kr}}

\begin{abstract}
We study possible new physics contributions in $B\to\pi\tau\bar\nu$
and $B\to\tau\bar\nu$ employing the model-independent effective 
Lagrangian that describes the quark-level transition 
$b\to u\tau\bar\nu$ at low energies.
The decay rate of $B\to\pi\tau\bar\nu$ and its theoretical uncertainty
are evaluated using the $B\to\pi$ form factors given by recent lattice 
QCD studies. Comparing theoretical results with the current experimental data,
$\mathcal{B}(B\to\pi\tau\bar\nu)<2.5\times 10^{-4}$ and
$\mathcal{B}(B\to\tau\bar\nu_\tau)=(1.14\pm 0.22)\times 10^{-4}$,
we obtain constraints on the Wilson coefficients that quantify 
potential new physics.
We also present the expected sensitivity of the SuperKEKB/Belle~II
experiment.
\end{abstract}


\maketitle

\section{Introduction}
\label{Sec:Intro}
Discrepancy of $\sim 4\sigma$ between experimental results and 
the standard model (SM) exists in the semitauonic $B$ meson decays,
$\bar B \to D^{(*)}\tau\bar\nu_\tau$ 
\cite{Lees:2012xj,Lees:2013uzd,Aaij:2015yra,Huschle:2015rga,Abdesselam:2016cgx}.
This anomaly is interesting apart from its statistical significance
in the sense that it suggests a manifestation of new physics
beyond the SM in the tree-level charged current SM processes 
involving the third-generation quark and lepton.

Since the interaction of quarks and leptons in the third generation might be
a clue to new physics, it is natural to search for a similar effect in the 
$b\to u\tau\bar\nu$ transition\footnote
{%
The charge-conjugated mode is implicit in the present work.
}. 
The evidence of the purely tauonic decay, 
$B^-\to\tau^-\bar\nu$, has been found by both the BaBar and Belle
collaborations and the combined value of their results of the branching
fraction is $\mathcal{B}(B^-\to\tau^-\bar\nu)=(1.14\pm 0.22)\times 10^{-4}$%
~\cite{Amhis:2014hma}, which is consistent with the SM prediction. 
Recently, the Belle collaboration reported on the semitauonic decay, 
$\bar B^0\to\pi^+\tau^-\bar\nu$~\cite{Hamer:2015jsa}. 
They observed no significant signal and obtained an upper limit of
the branching fraction as 
$\mathcal{B}(\bar B^0\to\pi^+\tau^-\bar\nu)<2.5\times 10^{-4}$
at the 90\% confidence level (CL).
As given in Ref.~\cite{Hamer:2015jsa}, 
the observed signal strength is $\mu = 1.52 \pm 0.72$, where $\mu =1$ 
corresponds to the branching fraction in units of $10^{-4}$, and thus one obtains 
\begin{equation}\label{Eq:BelleBPTN}
\mathcal{B}(\bar B^0\to\pi^+\tau^-\bar\nu)=
(1.52\pm 0.72\pm 0.13)\times 10^{-4}\,,
\end{equation}
where the second error comes from the systematic
uncertainty (8\%). Since the SM predicts $\sim 0.7 \times 10^{-4}$,
a new physics contribution of similar magnitude to the SM is allowed.
We expect that the SuperKEKB/Belle~II experiment will provide 
important information on possible new physics in 
$\bar B^0\to\pi^+\tau^-\bar\nu$ as well as $B^-\to\tau^-\bar\nu$.

Sensitivity to new physics effects depends on the precision of theoretical 
predictions as well as experimental errors.
The major uncertainty in the SM prediction of
$\mathcal{B}(\bar B^0\to\pi^+\tau^-\bar\nu)$ is ascribed to
the Cabibbo-Kobayashi-Maskawa matrix element $|V_{ub}|$ and 
the $B\to\pi$ hadronic form factors. In order to reduce these uncertainties,
it is useful to introduce the ratio of branching fractions
\cite{Chen:2006nua,Khodjamirian:2011ub,Bernlochner:2015mya},
\begin{align}\label{Eq:Rpi}
R_\pi=\frac{\mathcal{B}(\bar B^0\to\pi^+\tau^-\bar\nu)}
           {\mathcal{B}(\bar B^0\to\pi^+\ell^-\bar\nu) }\,,
\end{align}
as in the study of $\bar B \to D^{(*)} \tau\bar\nu_\tau$. Although
$|V_{ub}|$ cancels out in this ratio, there remains the uncertainty 
due to the form factors. Using the result of the recent lattice QCD study%
~\cite{Lattice:2015tia}, in which the relevant form factors are
obtained by fitting both the lattice amplitude and the experimental data 
of $B\to\pi\ell\bar\nu$%
~\cite{delAmoSanchez:2010af,Ha:2010rf,Lees:2012vv,Sibidanov:2013rkk},
the SM prediction is obtained as $R_\pi^\text{SM}=0.641\pm 0.016$%
~\cite{Bernlochner:2015mya,Du:2015tda}\footnote{
Ref.~\cite{Nandi:2016wlp} gives a different SM prediction. Our evaluation below agrees with Refs.~\cite{Bernlochner:2015mya,Du:2015tda}. 
}.
The experimental value is estimated as $R_\pi^\text{exp}\simeq 1.05\pm 0.51$,
where $\mathcal{B}(B\to\pi\ell\bar\nu)=(1.45\pm 0.02 \pm 0.04)\times 10^{-4}$%
~\cite{Amhis:2014hma} is used\footnote
{%
This is not the same way to obtain the experimental result of  
$R_{D^{(*)}}=\mathcal{B}(\bar B\to D^{(*)}\tau\bar\nu)
             /\mathcal{B}(\bar B\to D^{(*)} \ell\bar\nu)$%
~\cite{Lees:2012xj,Lees:2013uzd,Aaij:2015yra,Huschle:2015rga,
       Abdesselam:2016cgx}. 
The ratios $R_{D^{(*)}}$ are directly extracted with the signal events in 
the numerator and the normalization ones in the denominator
both involved in the same event sample.
}. 
New physics effects in $R_\pi$ and related quantities
are studied in the literature. 
The effect of charged Higgs boson, which appears
in the supersymmetric extension of the SM, is studied in 
Refs.~\cite{Chen:2006nua,Khodjamirian:2011ub,Bernlochner:2015mya}. 
The supersymmetric SM without $R$ parity is also studied in 
$b\to u$ (semi)leptonic processes~\cite{Kim:2007uq}.

In the present work, we study new physics effects in $B\to\pi\tau\bar\nu$ and $B\to\tau\bar\nu$ 
using the model-independent effective
Lagrangian that describes the $b\to u\tau\bar\nu$ transition at low energies.
Comparing with the current experimental data, we obtain constraints
on the Wilson coefficients that quantify potential new physics.
The theoretical uncertainties of $R_\pi$ in both the SM and new physics 
contributions are examined with the lattice QCD results. 
We also discuss prospects of new physics search in $B\to\pi\tau\bar\nu$ and
$B\to\tau\bar\nu$ at SuperKEKB/Belle~II.

This paper is organized as follows.  In \Sec{Sec:NPE}, we will
introduce the $b\to u\tau\bar\nu$ effective Lagrangian
that describes possible new physics contributions to 
$B \to (\pi) \tau\bar\nu$. We will also provide the relevant
rate formulae and theoretical uncertainties derived from errors of
form factor parameters given by lattice studies. 
In \Sec{Sec:NS}, we will present current constraints on new physics
from $B\to\pi\tau\bar\nu$ and $B\to\tau\bar\nu$, and 
discuss future prospects at SuperKEKB/Belle~II. A summary will be
given in \Sec{Sec:Summary}.

\section{Formulae of new physics effects}
\label{Sec:NPE}

\subsection{Effective Lagrangian}
In order to represent possible new physics effects at low energies, 
we adopt the model-independent approach with use of an effective Lagrangian~\cite{Tanaka:2012nw,Dutta:2013qaa}. 
As in our previous work~\cite{Tanaka:2012nw}, we assume that $b\to u\tau\bar\nu_\tau$ is affected by new physics while $b\to u\ell\bar\nu$ ($\ell = e, \mu$) is practically described by the SM. 
The effective Lagrangian used in this work is given by
\begin{equation}
\label{Eq:EL}
 -\mathcal{L}_\text{eff} = 2\sqrt{2}G_F V_{ub} 
 \Big[ 
 (1 +C_{V_1})\mathcal{O}_{V_1}+C_{V_2} \mathcal{O}_{V_2} +C_{S_1} \mathcal{O}_{S_1} +C_{S_2} \mathcal{O}_{S_2} +C_T \mathcal{O}_T
  \Big], 
\end{equation}
where the four-fermion operators are defined as   
\begin{align}
 & \mathcal{O}_{V_1} = ( \bar{u} \gamma^\mu P_L b)  (\bar{\tau} \gamma_\mu P_L \nu_\tau) \,, \label{Eq:OV1} \\
 & \mathcal{O}_{V_2} = ( \bar{u} \gamma^\mu P_R b)  (\bar{\tau} \gamma_\mu P_L \nu_\tau) \,,\\
 & \mathcal{O}_{S_1} = ( \bar{u} P_R b)  (\bar{\tau} P_L \nu_\tau) \,,\\ 
 & \mathcal{O}_{S_2} = ( \bar{u} P_L b)  (\bar{\tau} P_L \nu_\tau) \,,\\ 
 & \mathcal{O}_T       =  ( \bar{u} \sigma^{\mu\nu} P_L b)  (\bar{\tau} \sigma_{\mu\nu} P_L \nu_\tau) \,, \label{Eq:OT}
\end{align}
and $C_X$ ($X=V_{1,2},S_{1,2},T$) denotes the Wilson coefficient of
$\mathcal{O}_X$ normalized by $2\sqrt{2}G_F V_{ub}$.  
We only consider $\tau$-$\nu_\tau$ currents for simplicity though
the neutrino flavor could be the first or second generation in some new
physics models. One may translate the following result of $C_X$
for $\nu_{\ell = \tau}$ into that for $\nu_{\ell \neq \tau}$ by
replacing $C_X \to i |C_X|$.  Since $(\bar{u} \sigma^{\mu\nu} P_R b)
(\bar{\tau} \sigma_{\mu\nu} P_L \nu_\ell) =0$, there is only one
possible tensor operator unless right-handed neutrinos are
included in the low energy particle spectrum.
The SM contribution is represented by the unit coefficient of
$\mathcal{O}_{V_1}$, namely putting $C_X=0$ for all $X$'s gives the SM. 

In this paper, we focus on new physics effects in  
$B\to\pi\tau\bar\nu_\tau$ and $B\to\tau\bar\nu_\tau$. 
Other processes such as $B\to V\tau\bar\nu_\tau$ for $V = \rho, \omega$ might
become useful in future, but for now no experimental data are 
available. 

\subsection{$\bar B^0\to\pi^+\tau^-\bar\nu_\tau$}

The $B\to\pi$ transition caused by the effective Lagrangian in \eq{Eq:EL}
is described by the hadronic matrix elements of the quark currents involved
in the four-fermion operators:
\begin{align}
 \langle \pi (p_\pi) | \bar{u} \gamma^\mu b | \bar B(p_B) \rangle = 
 f_+(q^2) \left[ (p_B+p_\pi)^\mu - \frac{m_B^2 - m_\pi^2}{q^2} q^\mu
   \right] +f_0(q^2) \frac{m_B^2 - m_\pi^2}{q^2} q^\mu \,,
\end{align}
\begin{align}
 \langle \pi (p_\pi) | \bar{u} b | \bar B(p_B) \rangle = 
  (m_B+m_\pi)f_S(q^2)\,,
\end{align}
\begin{align}
 \langle \pi (p_\pi) | \bar{u} \, i \sigma^{\mu\nu} \, b | B(p_B) \rangle = \frac{2}{m_B + m_\pi} f_T (q^2) \left[ p_B^\mu p_\pi^\nu -  p_B^\nu p_\pi^\mu \right] \,,
\end{align}
where $q^\mu = (p_B - p_\pi)^\mu = (p_\tau + p_\nu)^\mu$, and
$f_{+,0,S,T}(q^2)$ are form factors.  
We note that the axial-vector (pseudoscalar) part of $V_{1,2}$ ($S_{1,2}$), 
$\bar{u} \gamma^\mu\gamma^5 b$ ($\bar{u}\gamma^5 b$),
does not contribute to the transition, and 
$\langle \pi(p_\pi) | \bar{u} \sigma^{\mu\nu} \gamma^5 b | B(p_B) \rangle$
is expressed by $f_T (q^2)$ with 
$\sigma^{\mu\nu}\gamma^5 = -\frac{i}{2} \varepsilon^{\mu\nu\alpha\beta}
 \sigma_{\alpha\beta}$\footnote{  
We take $\varepsilon^{0123}=-1$.}.   
We employ the vector and tensor form factors $f_{+,0,T}$ given by
recent lattice QCD studies~\cite{Lattice:2015tia,Bailey:2015nbd}.
As for the scalar form factor $f_S$, since no lattice evaluation is available at present, we utilize the quark 
equation of motion to relate $f_S$ to 
$f_0$, namely $f_S(q^2)=f_0(q^2)(m_B-m_\pi)/(m_b-m_u)$.

The differential branching fractions of $B \to \pi\tau\bar\nu_\tau$ for 
given $\tau$ helicities, defined in the rest frame of the lepton pair, are written as
\begin{align}
 \frac{d \mathcal B_\tau^-}{d q^2} 
=
 N_B  \Big | (1 +C_{V_1} +C_{V_2}) \sqrt{q^2} H_{V_+} +4 C_T m_\tau H_T \Big |^2 \,,\label{Eq:BM}
\end{align}
for $\lambda_\tau = -1/2$, and
\begin{align}
 \frac{d \mathcal B_\tau^+}{d q^2} 
 = \frac{N_B}{2} \bigg[ 
 & \Big| (1+C_{V_1}+C_{V_2}) m_\tau H_{V_+} +4 C_T \sqrt{q^2} H_T \Big|^2 \notag \\
 & +3 \Big| (1+C_{V_1}+C_{V_2}) m_\tau H_{V_0} +(C_{S_1}+C_{S_2}) \sqrt{q^2} H_{S} \Big|^2 \bigg] \,,\label{Eq:BP}
\end{align}
for $\lambda_\tau = +1/2$, with
\begin{align}
 N_B = 
 \frac{\tau_{B^0} G_F^2 |V_{ub}|^2}{192 \pi^3 m_B^3} \sqrt{Q_+ Q_-} \left( 1 -\frac{m_\tau^2}{q^2} \right)^2 \,,
\end{align}
where $\tau_{B^0}$ is the neutral $B$ meson lifetime and 
$Q_\pm = (m_B \pm m_\pi)^2 - q^2$. 
The hadronic amplitudes $H$'s are given by 
\begin{align}
 H_{V_+} &= \frac{\sqrt{Q_+ Q_-}}{\sqrt{q^2}} f_+(q^2) \,, \label{Eq:HV}\\
 H_{V_0} &= \frac{m_B^2 - m_\pi^2}{\sqrt{q^2}} f_0(q^2) \,, \\
 H_S &=(m_B+m_\pi)f_S(q^2)=\frac{m_B^2 - m_\pi^2}{m_b - m_u} f_0 (q^2) \,, \\
 H_T &= \frac{\sqrt{Q_+ Q_-}}{m_B + m_\pi} f_T (q^2) \,, 
\end{align}
where the bottom and up quark masses are taken as 
$m_b=4.2\ \text{GeV}$ and $m_u = 0$ in the following numerical calculation.
The differential branching fractions of $B \to \pi \ell\bar\nu_\ell$ 
(for $m_\ell=0$) are obtained as
\begin{align}
 & \label{Eq:bpiellnum} \frac{d \mathcal B_\ell^-}{d q^2} = \left . \frac{d \mathcal B_\tau^-}{d q^2} \right|_{m_\tau \to 0,\, C_X = 0}  \,, \\
 & \label{Eq:bpiellnup} \frac{d \mathcal B_\ell^+}{d q^2} = 0 \,. 
\end{align}
In the following, the ratio of the branching fractions, 
$R_\pi$ in Eq.~\eqref{Eq:Rpi}, is numerically calculated by 
\begin{align}
 R_\pi 
 = \frac{\displaystyle \int_{m_\tau^2}^{(m_B + m_\pi)^2} dq^2~\frac{d \mathcal B_\tau^+ + d \mathcal B_\tau^-}{dq^2} }
 {\displaystyle \int_0^{(m_B + m_\pi)^2} dq^2~\frac{d \mathcal B_\ell^-}{dq^2} } \,. 
\end{align} 
As mentioned above, $|V_{ub}|$ cancels out in this ratio, but
errors in the form factors cause the theoretical uncertainty in $R_\pi$.

The form factors $f_+$, $f_0$ and $f_T$ are parametrized with the use
of the Bourrely-Caprini-Lellouch expansion as%
~\cite{Lattice:2015tia,Bailey:2015nbd,Bourrely:2008za}
\begin{align}
 & f_j (q^2) = \frac{1}{1-q^2/m_{B^*}^2} \sum_{n=0}^{N_z-1} b_n^j \left[ z^n - (-1)^{n - N_z} \frac{n}{N_z} z^{N_z} \right] \,, \\
 & f_0 (q^2) = \sum_{n=0}^{N_z-1} b_n^0 z^n \,,
\end{align}
where $j=+,T$, $m_{B^*} = 5.325\,\text{GeV}$ is the $B^*$ meson mass,
$b_n^{+,0,T}$ are expansion coefficients, and $N_z=4$ is the expansion
order.  The expansion parameter $z$ is defined as
\begin{align}
 z \equiv z(q^2) = \frac{ \sqrt{t_+ - q^2} - \sqrt{t_+ - t_0} }{\sqrt{ t_+ - q^2} + \sqrt{t_+ - t_0} } \,,
\end{align}
where $t_+ =(m_B + m_\pi)^2$ and 
$t_0 = (m_B + m_\pi) (\sqrt{m_B} - \sqrt{m_\pi})^2$. 
The combined fit to the experimental data of  
the $q^2$ distribution of $B \to\pi\ell\bar\nu_\ell$ and the lattice computation for the relevant amplitudes
provides the ``lattice+experiments'' fitted values of 
$b_n^{+,0,T}$ with errors and their correlations. According to
Refs.~\cite{Lattice:2015tia,Bailey:2015nbd}, the result of
the expansion coefficients 
$\vec{b} = (b^+_0, b^+_1, b^+_2, b^+_3, b^0_0, b^0_1, b^0_2, b^0_3, b^T_0, b^T_1, b^T_2, b^T_3)^\intercal$ is summarized as
\begin{align}
 \vec{b}_\text{lat.+exp.} \equiv \vec{b}_0 \pm \delta \vec{b} \,,
\end{align}
where 
\begin{align}
 \label{Eq:expara_cent}
 \hspace{-3em} \vec{b}_0 =
 (
  0.419,-\!0.495,-\!0.43, 0.22, 0.510, -\!1.700, 1.53, 4.52, 0.393, -\!0.65, -\!0.6, 0.1 
 )^\intercal \,,
\end{align}
\begin{align}
 \label{Eq:expara_err}
 \delta \vec{b} =
 (
  0.013, 0.054, 0.13, 0.31, 0.019, 0.082, 0.19, 0.83, 0.017, 0.23, 1.5, 2.8 
 )^\intercal \,.  
\end{align}
We note that only $b_n^+$'s are directly constrained by the experimental
data because only $f_+(q^2)$ contributes to $B \to\pi\ell\bar\nu_\ell$
as seen in Eqs.~\eqref{Eq:BM}, \eqref{Eq:HV}, \eqref{Eq:bpiellnum} and 
\eqref{Eq:bpiellnup}. In addition, $b_n^0$'s are indirectly constrained
through the relation $f_0(0)=f_+(0)$. The tensor form factor $f_T(q^2)$
is determined thoroughly by the lattice simulation and this explains 
the relatively large errors of $b_n^T$'s. 

\textcolor{black}{
The covariance matrix is} given by 
$V_{ij}= \rho_{ij} \delta b_i \delta b_j$ with 
\begin{align}
 \rho_\text{lat.+exp.} = 
 \begin{pmatrix}
  \rho_{+,0} & {\bm 0}_{8 \times 4} \\
 {\bm 0}_{4 \times 8} & \rho_{T}  
 \end{pmatrix} \,,
\end{align}
\begin{align}
 \rho_{+,0} = 
 \begin{pmatrix}
 1 &		0.14 	&	-0.455 &		-0.342 &		0.224 	&		0.174  &		0.047 &		-0.033 \\ 
    &		1 	&	-0.789 &		-0.874 &		-0.068 	&		0.142  &		0.025 &		-0.007 \\
    &			&		1 &		0.879  &		-0.051	&		-0.253 &		0.098 &		0.234 \\
    &			&		   &		1	   &		0.076	&		0.038  &		0.018 &		-0.2 \\
    &			&		   &			   &		1		&		-0.043 &		-0.604&		-0.388 \\
    &			&		   &			   &				&		1	   &		-0.408&		-0.758 \\
    &			&		   &			   &				&			   &		1	  &		0.457 \\
    &			&		   &			   &				&			   &			  &		1 
 \end{pmatrix} \,,
\end{align}
\begin{align}
 \rho_{T} = 
 \begin{pmatrix}
 1	&	0.4	&	0.204	&	0.166	\\
 	&	1	&	0.862	&	0.806	\\
	&		&	1		&	0.989	\\
	&		&			&	1
 \end{pmatrix} \,,
\end{align}
where $\rho$'s are symmetric correlation matrices.  Here, we have 
omitted the correlations between the $+,0$ sector and the $T$ 
sector, because the covariance matrix turns out not to be positive
semidefinite if all the correlations reported in 
Refs.~\cite{Lattice:2015tia,Bailey:2015nbd} are taken.
Negative eigenvalues of a covariance matrix may arise due to the fluctuation of eigenvalues. 
In such a case, the correlation is less significant and could be neglected. 

The error of $\vec{b}$ induces the uncertainty in both the SM and 
new physics contributions in the observable $R_\pi$. 
To estimate the uncertainty of $R_\pi$, we calculate its variance 
$V(R_\pi)$ assuming the Gaussian distribution:
\begin{align}
 \label{Eq:RpiError}
 V(R_\pi) = \int d \vec{b} \left( R_\pi (\vec{b}) - R_\pi (\vec{b}_0) \right)^2 \exp \left[ -\frac{1}{2} \chi^2 ( \vec{b}) \right] \,,
\end{align}
\begin{align}
 \chi^2 ( \vec{b}) = \left( \vec{b} - \vec{b}_0 \right)^\intercal V(\vec{b})^{-1} \left( \vec{b} - \vec{b}_0 \right) \,. 
\end{align}
The theoretical uncertainty of $R_\pi$ is thus given by 
$\delta R_\pi =\sqrt{V(R_\pi)}$. 

\subsection{$B^- \to \tau^-\bar\nu_\tau$}
The branching fraction of $B^-\to\tau^-\bar\nu_\tau$ in the effective 
Lagrangian in \eq{Eq:EL} is expressed as
\begin{align}
 \label{Eq:Btaunu}
 \mathcal B (B \to \tau\bar\nu_\tau) = \frac{\tau_{B^-} G_F^2 |V_{ub}|^2 f_B^2}{8\pi} m_B m_\tau^2 \left( 1 - \frac{m_\tau^2}{m_B^2} \right)^2 \left| 1 + r_\text{NP} \right|^2 \,,
\end{align}
where $\tau_{B^-}$ is the charged $B$ meson lifetime, $f_B$ is the $B$ meson decay constant, and $r_\text{NP}$ represents the new physics effect, 
\begin{align}
 \label{Eq:rNP}
 r_\text{NP} = C_{V_1} - C_{V_2} + \frac{m_B^2}{m_b m_\tau} \left( C_{S_1} - C_{S_2} \right) \,.  
\end{align} 
We note that the tensor operator $\mathcal O_T$ does not contribute to this decay mode.

The dominant sources of theoretical uncertainty in $\mathcal{B}(B\to\tau\bar\nu_\tau)$ are $f_B$ and $|V_{ub}|$.  
The FLAG working group gives an average of lattice QCD results~\cite{McNeile:2011ng,Bazavov:2011aa,Na:2012kp,Christ:2014uea,Aoki:2014nga} as $f_B = (192.0 \pm 4.3)\ \text{MeV}$~\cite{Aoki:2016frl}, which is consistent with another average~\cite{Rosner:2015wva}. 
As for $|V_{ub}|$, the tension among the values determined from $B \to \pi\ell\bar\nu_\ell$ (exclusive), $B \to X_u\ell\bar\nu_\ell$ (inclusive) and the fit of the unitarity triangle is still unsolved.
To avoid the uncertainty due to $|V_{ub}|$, the following ratio of pure- and semi- leptonic decay rates is defined as~\cite{Fajfer:2012jt}
\begin{align}
 R_\text{ps}  
 = \frac{ \Gamma (B^- \to \tau^-\bar\nu_\tau) }{ \Gamma (\bar B^0 \to \pi^+ \ell^-\bar\nu_\ell)}
 = \frac{\tau_{B^0}}{\tau_{B^-}} \frac{ \mathcal B (B^- \to \tau^-\bar\nu_\tau) }{ \mathcal B (\bar B^0 \to \pi^+ \ell^-\bar\nu_\ell)} \,. 
\end{align}
The remaining sources of theoretical uncertainty in $R_\text{ps}$ are $f_B$ and the form factor $f_+(q^2)$ involved in the denominator. 
For the latter, we use the lattice result described above.

\section{Numerical results}
\label{Sec:NS}

\subsection{New physics scenarios}
\label{SubSec:NPS}

We consider new physics scenarios such that only one of the operators 
$\mathcal O_X$ ($X=V_1,V_2,S_1,S_2,T$) is dominant in the new physics sector.
These scenarios are constrained by both $B \to \pi\tau\bar\nu_\tau$ and 
$B \to \tau\bar\nu_\tau$ except the tensor operator scenario, in which
$B \to \tau\bar\nu_\tau$ is not altered.

First, we present numerical formulae of the theoretical uncertainties
$\delta R_\pi$ obtained by computing the variance in \eq{Eq:RpiError} 
for each scenario: 
\begin{align}
 & \label{Eq:ThErrorVi} \delta R_\pi (C_{V_i}, C_{X \neq V_i}=0) 
 \simeq \delta R_\pi^\text{SM} \left| 1 + C_{V_i} \right|^2 \,, \\[0.5em]
 & \label{Eq:ThErrorSi} \delta R_\pi (C_{S_i}, C_{X \neq S_i}=0) 
 \simeq \delta R_\pi^\text{SM} \,\Big( 1 + 7\, \ReC{S_i} + 15\, \ReC{S_i}^2 + 9\,\AbsC{S_i}^2   \notag \\
 & ~~~~~~~~~~~~~~~~~~~~~~~~~~~~~~~~~~~~~~~~~~+ 35\, \ReC{S_i} \AbsC{S_i}^2 +21\, \AbsC{S_i}^4 \Big)^{1/2} \,, \\[0.5em]
 & \label{Eq:ThErrorT} \delta R_\pi (C_{T}, C_{X \neq T}=0) 
 \simeq \delta R_\pi^\text{SM} \,\Big( 1+ 4\, \ReC{T} + 350\, \ReC{T}^2 + 11\, \AbsC{T}^2  \notag \\
 & ~~~~~~~~~~~~~~~~~~~~~~~~~~~~~~~~~~~~~~~~~~+ 1372\, \ReC{T} \AbsC{T}^2 + 1484\, \AbsC{T}^4 \Big)^{1/2} \,,
\end{align}
where $\delta R_\pi^\text{SM}\simeq 0.016$
represents the uncertainty in the SM,
which is consistent with the value in Refs.~\cite{Bernlochner:2015mya,Du:2015tda}. 
We observe that the contribution of the tensor operator is rather
uncertain because of the less-determined form factor $f_T (q^2)$
as mentioned above.

\begin{figure}[t!]
\begin{center}
\includegraphics[viewport=0 0 450 470, width=12em]{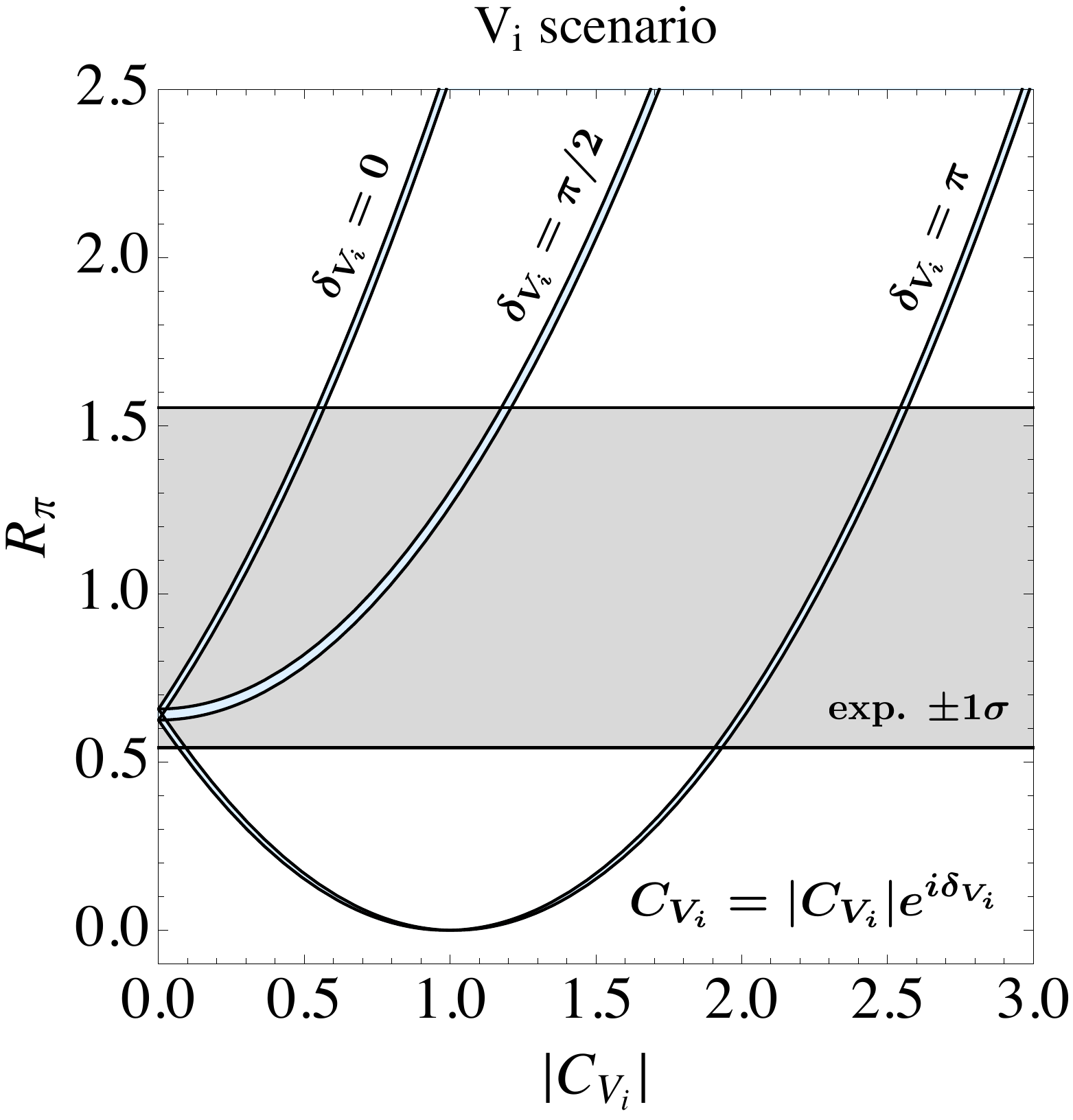}~~~ 
\includegraphics[viewport=0 0 450 470, width=12em]{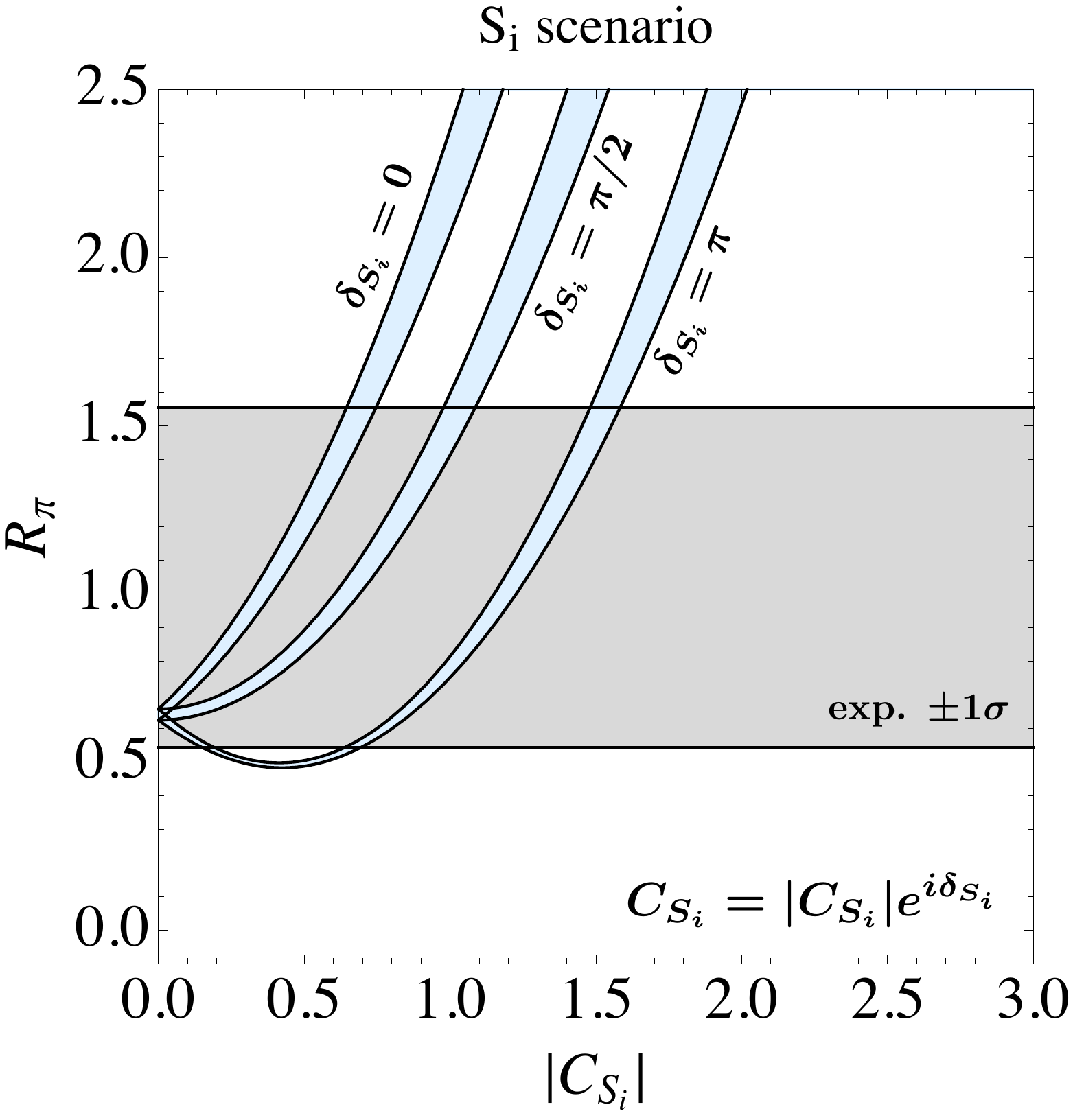}~~~ 
\includegraphics[viewport=0 0 450 470, width=12em]{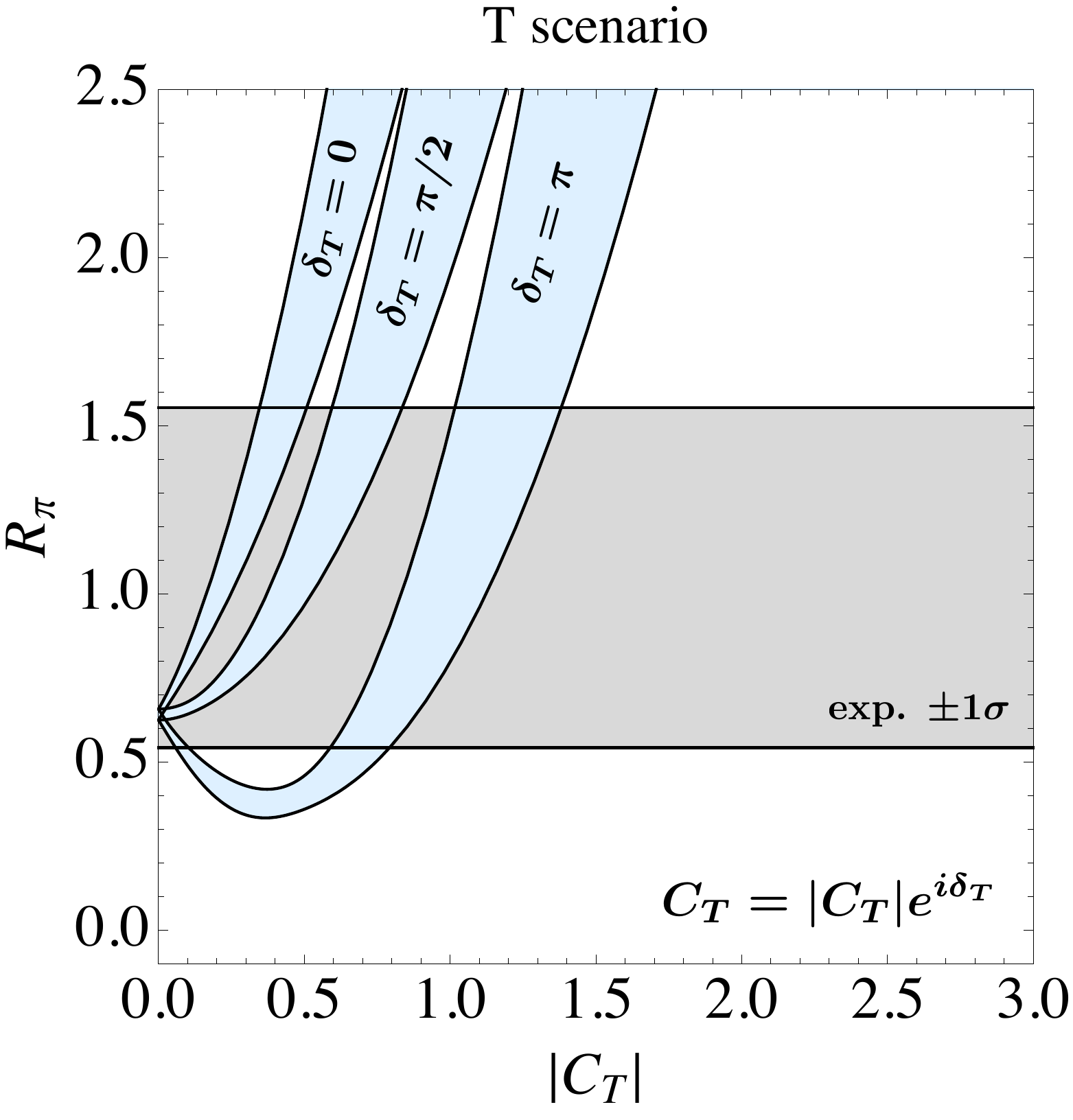} 
\caption{ 
  New physics effects on $R_\pi$ in the $V_i$, $S_i$, and $T$
  scenarios. Three values of the complex phase, 
  $\delta_X = 0$, $\pi/2$ and $\pi$, are chosen.
  The blue regions represent the theoretical predictions on $R_\pi$ 
  taking the theoretical uncertainty ($\pm 1\sigma$) into account.
  The gray regions show the current experimental bound, 
  $R_\pi^\text{exp.} \simeq 1.05 \pm 0.51$.}
\label{Fig:Effect}
\end{center}
\end{figure}
In Fig.~\ref{Fig:Effect}, we show $R_\pi$ in our new physics scenarios 
as functions of $|C_X|$ for three representative values of the complex
phase (defined by $C_X = |C_X| e^{i\delta_X}$) as indicated.  
The light blue regions are the theoretical predictions
with the $\pm 1 \sigma$ uncertainties evaluated with
Eqs.~(\ref{Eq:ThErrorVi})-(\ref{Eq:ThErrorT}).  The gray region
expresses the present experimental bound at the $1\sigma$ level as is
estimated in \Sec{Sec:Intro}.  One finds that the theoretical
uncertainty in the vector scenarios is fairly small compared with
the experimental error, whereas that in the tensor scenario is significant\footnote
{%
The uncertainty in the bottom quark mass, which is fixed in the present work,
increases the theoretical uncertainties in the scalar scenarios. Varying
$m_b$ by $\pm 200$ MeV changes at most $\mathcal{B}(B\to\pi\tau\bar\nu_\tau)$ and 
$\mathcal{B}(B\to\tau\bar\nu_\tau)$ by $\pm 6\%$ and $\pm 7\%$
respectively for $|C_{S_i}|<1$. 
}.

One may observe in Eqs.~\eqref{Eq:BM}, \eqref{Eq:BP} and \eqref{Eq:rNP}
that $\mathcal O_{V_1}$ and $\mathcal O_{V_2}$ 
have the same contribution to $B \to \pi\tau\bar\nu_\tau$ whereas 
their contributions to $B \to\tau\bar\nu_\tau$ posses opposite 
sign with each other. This is simply because the (axial-)vector
current $\bar u \gamma^\mu b$ ($\bar u \gamma^\mu \gamma^5 b$) 
contributes only to $B \to \pi$ transition ($B$ annihilation).
Thus, the vector and the axial vector parts of new physics,
namely $C_V=(C_{V_1}+C_{V_2})/2$ and $C_A=(C_{V_2}-C_{V_1})/2$ are 
separately constrained by $B \to \pi\tau\bar\nu_\tau$
and $B \to\tau\bar\nu_\tau$ respectively. 
The same argument applies to $\mathcal O_{S_1}$ and 
$\mathcal O_{S_2}$. The (pseudo)scalar part,
$C_S =(C_{S_1} + C_{S_2})/2$ ($C_P =(C_{S_1} - C_{S_2})/2$)
is constrained by $B \to \pi\tau\bar\nu_\tau$
($B \to\tau\bar\nu_\tau$). As stressed above,
the tensor operator $\mathcal O_{T}$ contributes only to 
$B \to\pi\tau\bar\nu_\tau$.

\subsection{Present constraints}

\begin{figure}[t!]
\begin{center}
\includegraphics[viewport=0 0 450 479, width=14em]{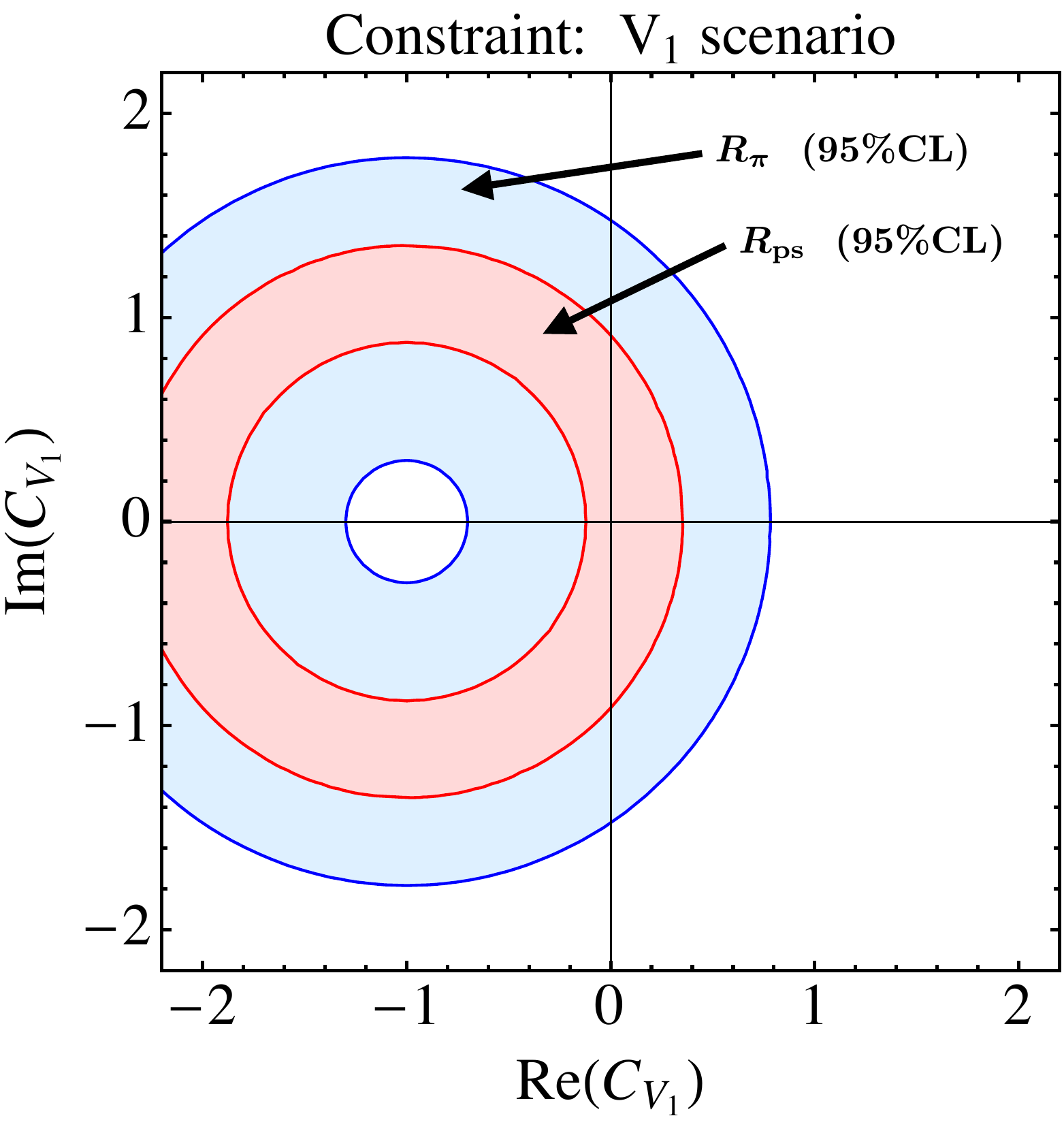} \hspace{2em}
\includegraphics[viewport=0 0 450 479, width=14em]{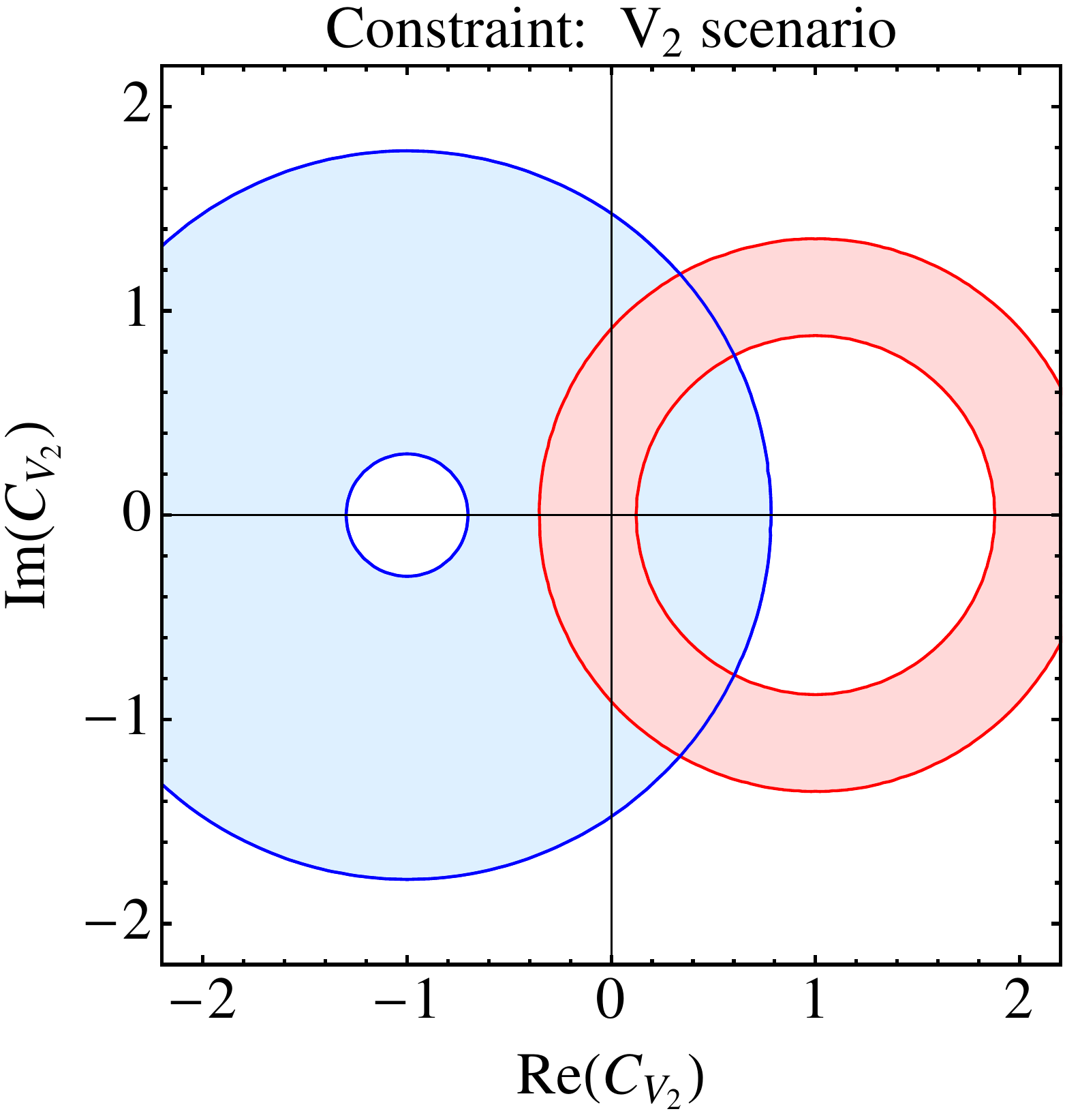} \\[2em]
\includegraphics[viewport=0 0 450 479, width=14em]{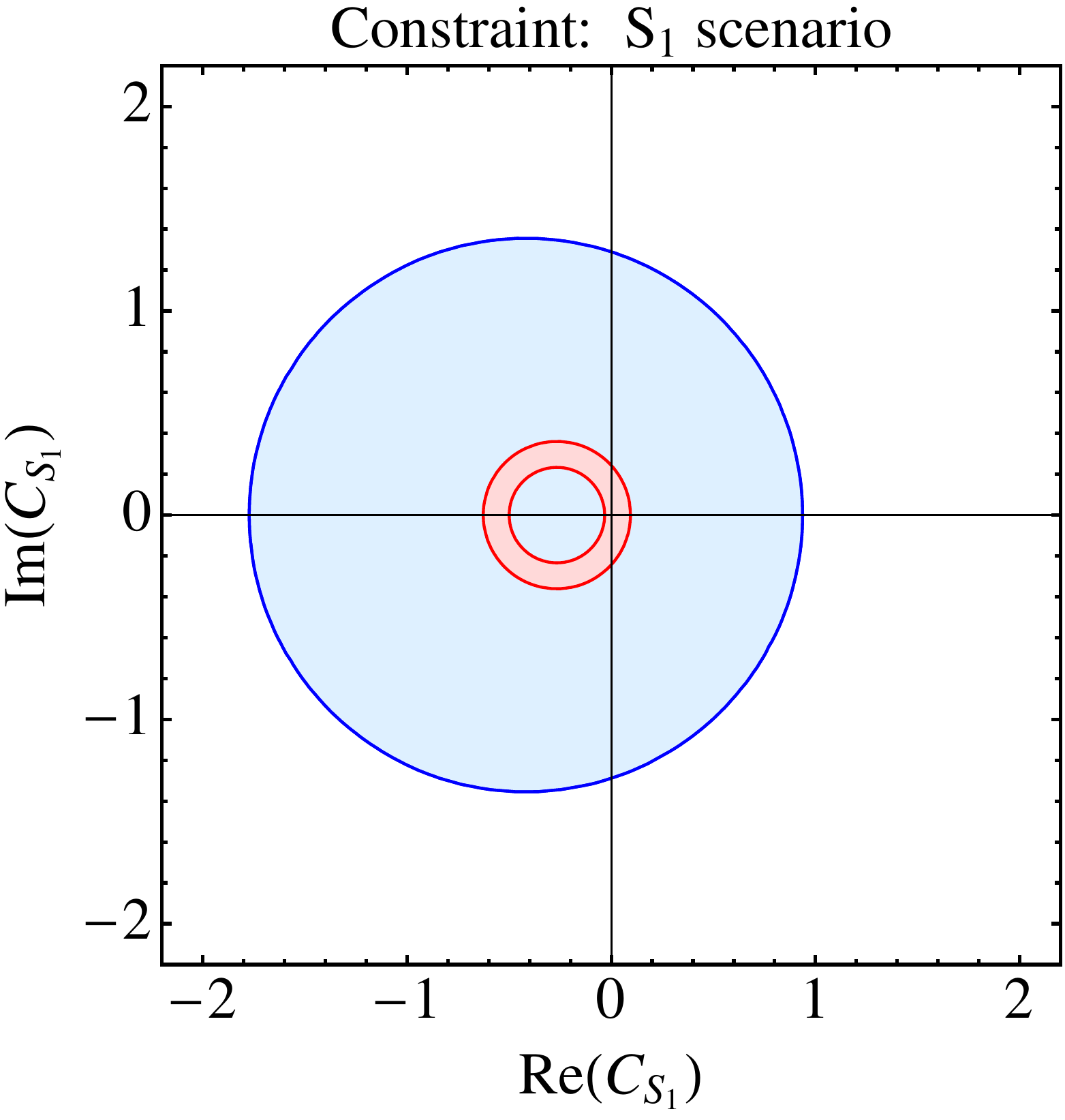} \hspace{2em}
\includegraphics[viewport=0 0 450 479, width=14em]{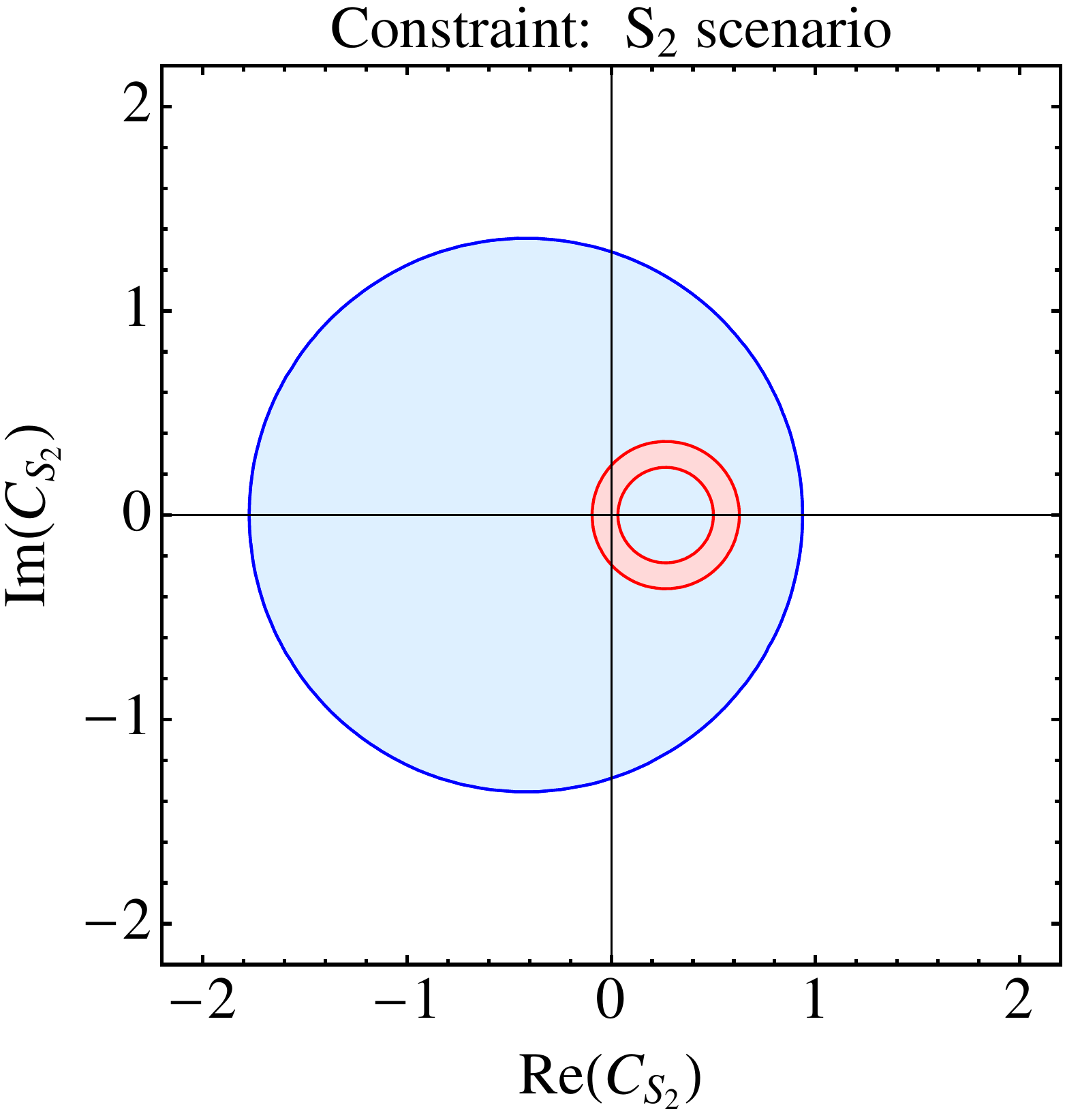} 
\caption{
Allowed regions from $R_\pi$ and $R_\text{ps}$ for $V_1$, $V_2$, $S_1$, and $S_2$ scenarios. 
The light blue region is allowed from the value of $R_\pi$ derived from the Belle experiment at 95\% CL, where the theoretical uncertainty is also taken for the evaluation. 
The light red region is consistent with the experimental value of $R_\text{ps}$ taking into account the theoretical uncertainty described in the main text. 
} 
\label{Fig:Bound}
\end{center}
\end{figure}
%
The current experimental result for $R_\pi$ is given in \Sec{Sec:Intro}, $R_\pi^\text{exp.} \simeq 1.05 \pm 0.51$. 
As for $R_\text{ps}$, we obtain $R_\text{ps}^\text{exp.}=0.73 \pm 0.14$, while the SM prediction is $R_\text{ps}^\text{SM} = 0.574 \pm 0.046$ including the uncertainties of $f_B$ and $f_+(q^2)$. 
Given these experimental data, 
we present constraints on the Wilson coefficients $C_X$'s for $V_1$, $V_2$, $S_1$, and $S_2$ scenarios in
Fig.~\ref{Fig:Bound}, in which the 95\% CL allowed regions by $R_\pi$ and $R_\text{ps}$ for each scenario are shown.  
The light blue and red regions are allowed by $R_\pi$ and $R_\text{ps}$, respectively, taking both 
the theoretical and experimental uncertainties into account.

The current data of $R_\pi$ excludes part of the region of 
$|C_X| \sim O(1)$, which is roughly the same order of magnitude as
the SM contribution.
The excluded region by $R_\pi$ does not exceed the one by 
$R_\text{ps}$ in the $V_1$ scenario, 
but their difference is not so significant. 
As for the $V_2$ scenario, $R_\pi$ and $R_\text{ps}$ 
are complementary because the signs of the new physics contributions
relative to the SM ones are opposite in these observables as seen in 
Eqs.~\eqref{Eq:BM}, \eqref{Eq:BP} and \eqref{Eq:rNP}.
The $S_1$ and $S_2$ scenarios are constrained more tightly by 
$R_\text{ps}$ because of the chiral enhancement 
of the pseudoscalar contribution in the purely leptonic decay.

\begin{figure}[t!]
\begin{center}
\includegraphics[viewport=0 0 450 479, width=18em]{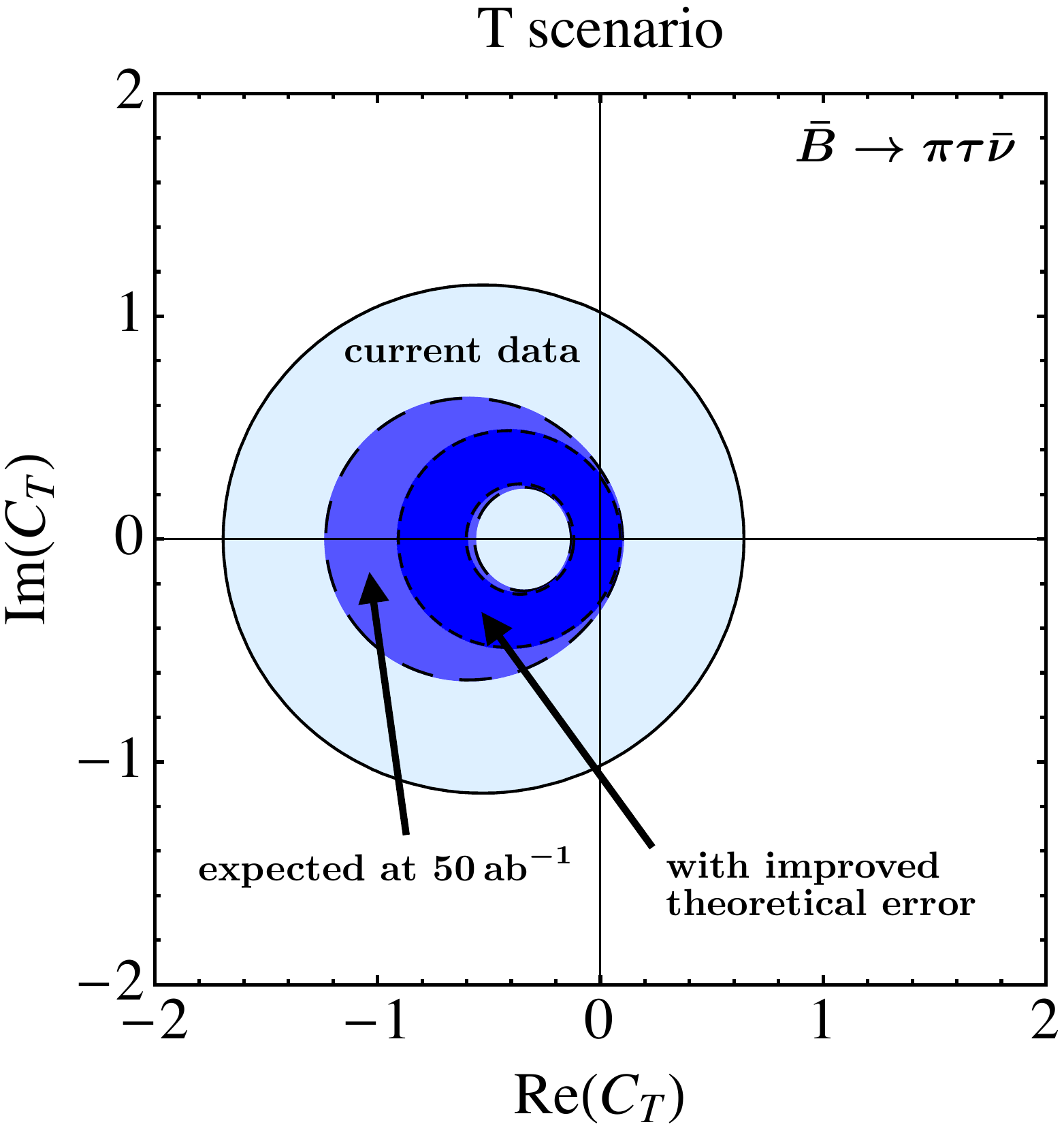} 
\caption{
Allowed regions of the tensor scenario for the recent experimental data of
$R_\pi$ and expected improvements in future. 
The present allowed region at the 95\% CL is depicted in light blue. 
The (dark) blue region enclosed by the dashed (dotted) curves shows 
the allowed region expected at SuperKEKB/Belle~II with 
$50 \,\text{ab}^{-1}$ data (and a theoretical uncertainty reduced by
a factor of 2). 
} 
\label{Fig:BoundT}
\end{center}
\end{figure}
%
In Fig.~\ref{Fig:BoundT}, we show the $R_\pi$ constraint on $C_T$.
The light blue region represents the 95\% CL allowed region.
(The darker blue regions will be explained below.)
We see that the present constraint is nontrivial and comparable to
the other scenarios even though the theoretical uncertainty are 
considerably larger. This is because the tensor operator (as 
normalized in Eq.~\eqref{Eq:EL}) tends to give a larger contribution
to $R_\pi$. The tensor contribution is expected to be more significant
in the $B\to V$ transitions, such as $\bar B \to\rho$ as
in the case of $\bar B\to D^*\tau\bar\nu_\tau$.

\subsection{Future prospect}

From now on, we discuss expected data of the relevant observables at SuperKEKB/Belle~II and estimate the possible sensitivity to the new physics scenarios. 
The current experimental value of $R_\pi$ given in Eq.~\eqref{Eq:BelleBPTN} is obtained with $\sim 1\ \text{ab}^{-1}$ data.
We expect $\sim 50\ \text{ab}^{-1}$ at the SuperKEKB/Belle~II experiment.
To evaluate the expected sensitivity of $R_\pi$ to the new physics scenarios at SuperKEKB/Belle~II, 
we assume that both the statistical and systematic errors in the experiment are reduced with increasing luminosity as $1/\sqrt{\mathcal L}$ and that the central value coincides with the SM prediction.  
Namely, we employ $R_\pi^\text{Belle II}=0.641\pm 0.071$. 
Applying a similar argument to $B\to \tau\nu$ and $B \to \pi\ell\nu$ gives $R_\text{ps}^\text{Belle II} = 0.574 \pm 0.020 $\footnote
{%
The expected Belle II sensitivity for $\mathcal B (B^-\to\tau^-\bar\nu)$ has been recently studied in Ref.~\cite{B2sensitivity:Btaunu} with Monte-Carlo simulation assuming $0.5\,\text{ab}^{-1}$. 
Our estimation and their result of the experimental uncertainty scaled to at $50\,\text{ab}^{-1}$ are consistent. 
}.

\begin{figure}[t!]
\begin{center}
\includegraphics[viewport=0 0 450 473, width=14em]{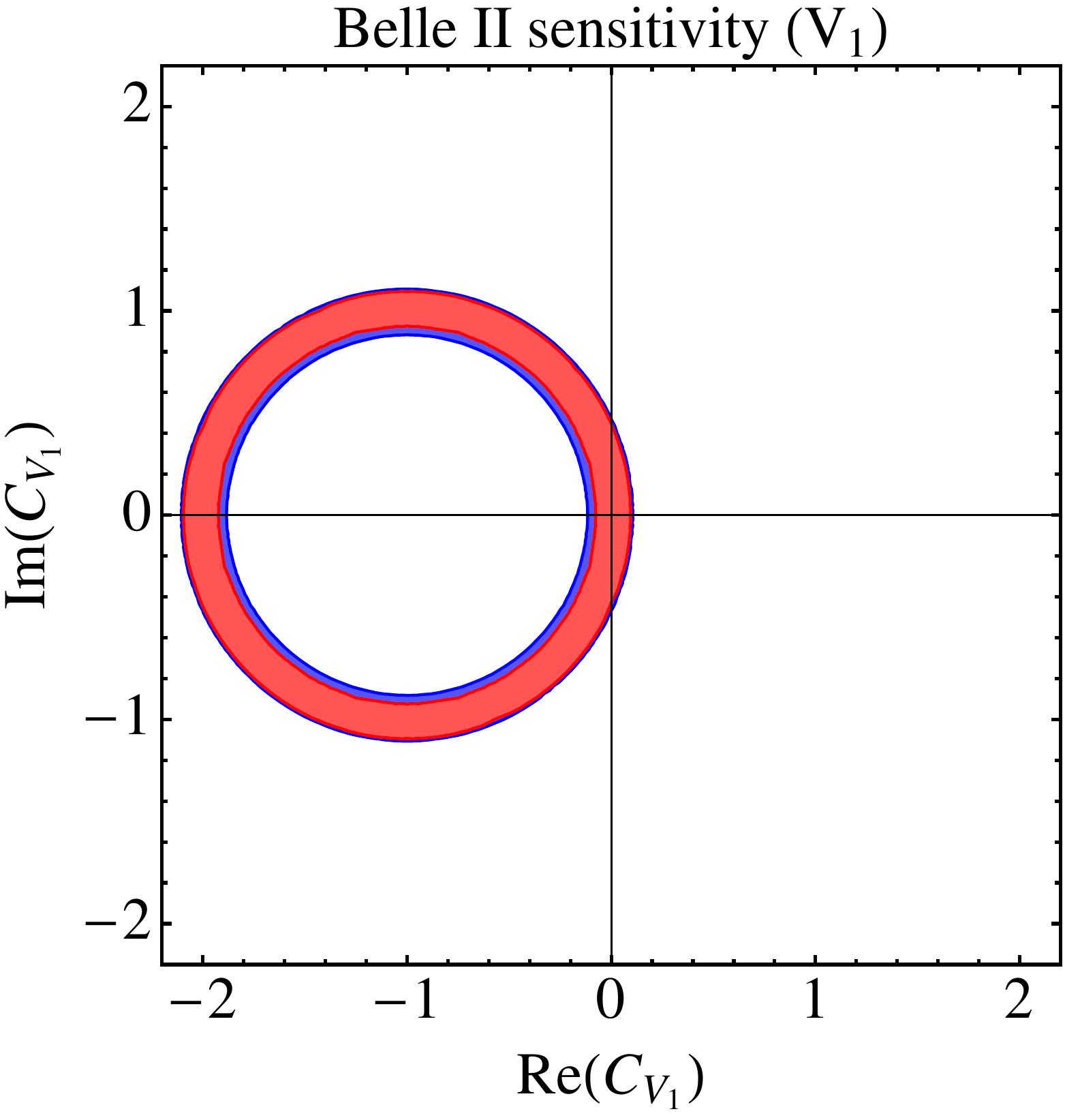} \hspace{2em}
\includegraphics[viewport=0 0 450 473, width=14em]{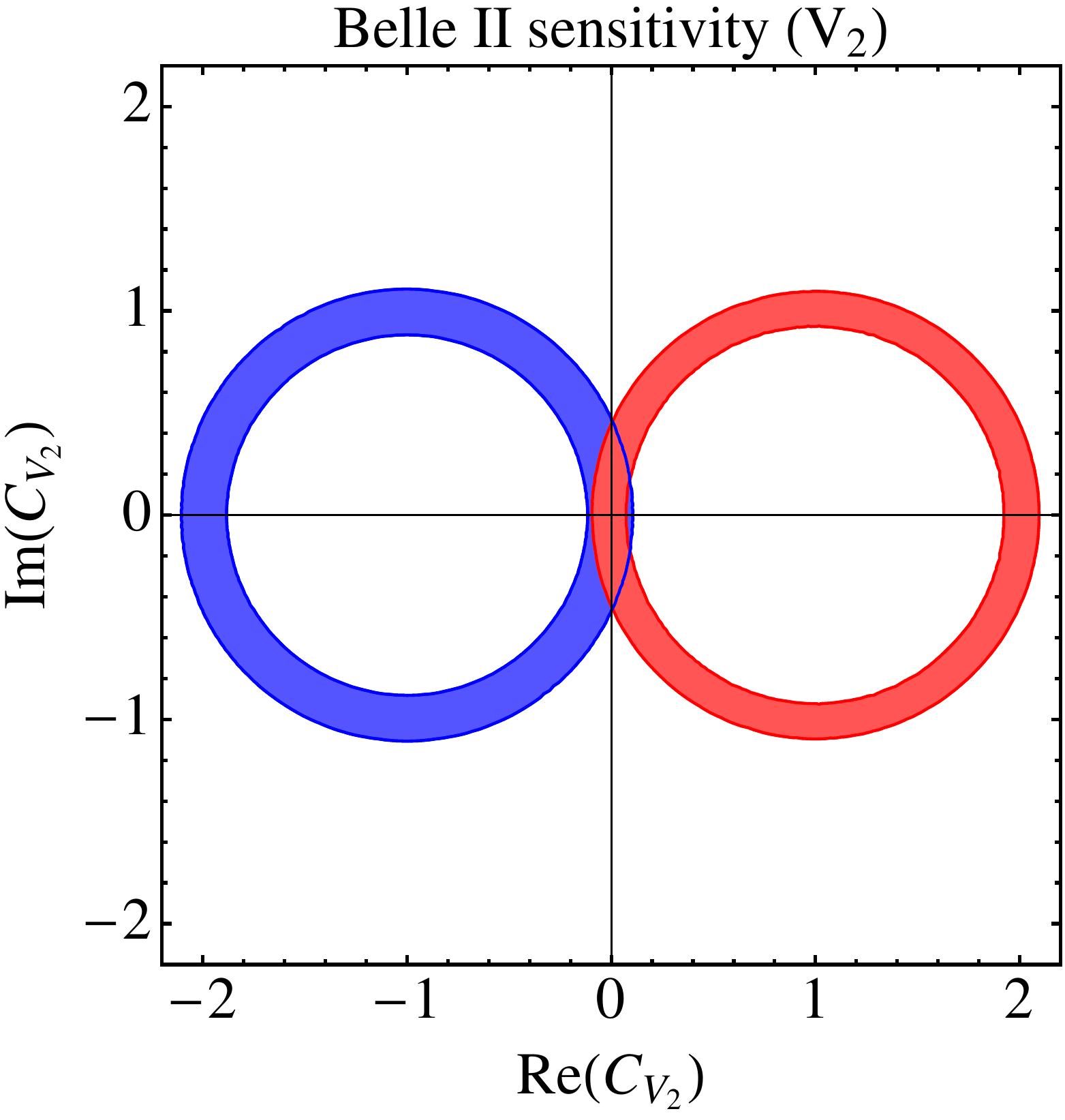} \\[2em]
\includegraphics[viewport=0 0 450 467, width=14em]{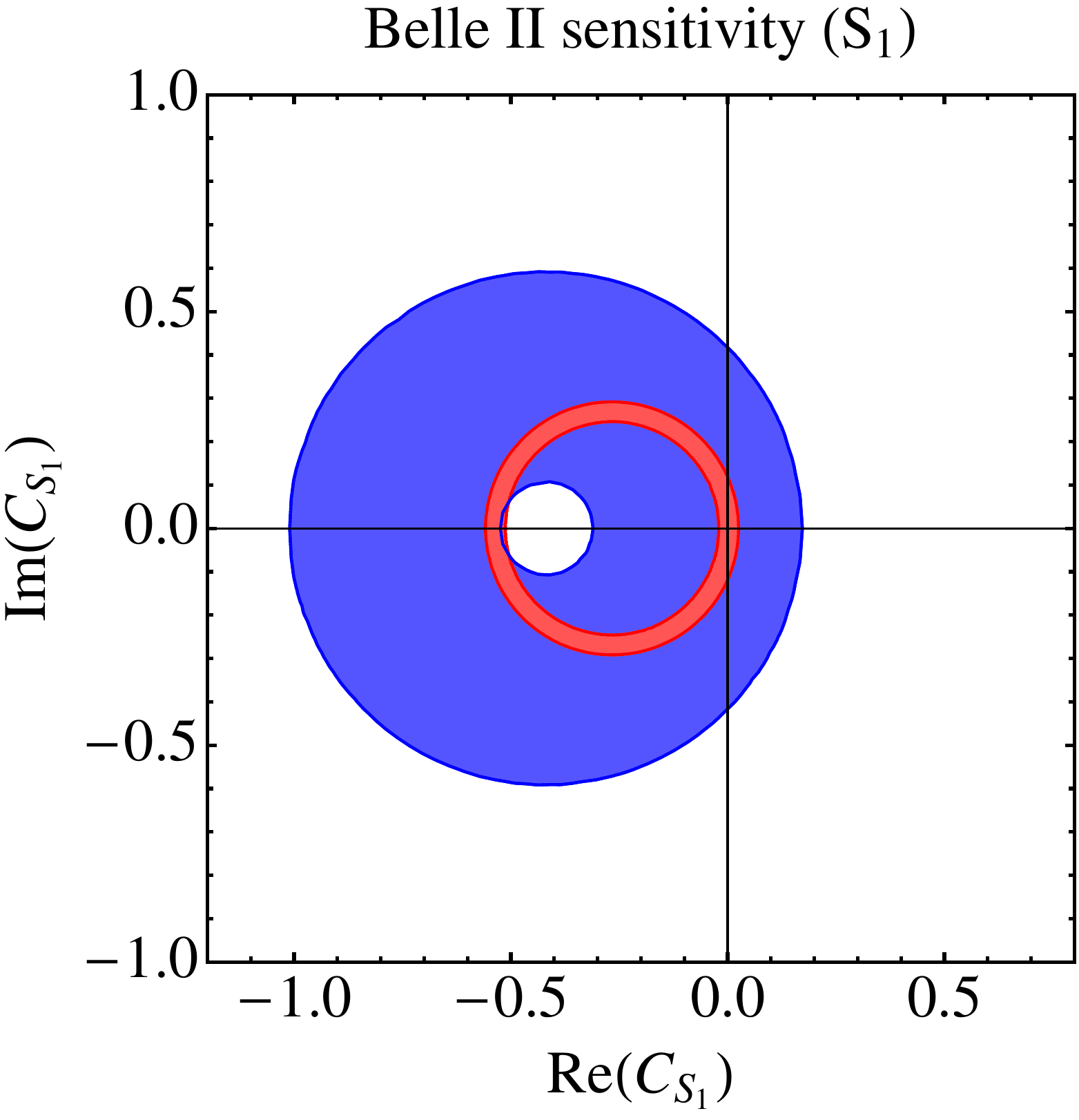} \hspace{2em}
\includegraphics[viewport=0 0 450 467, width=14em]{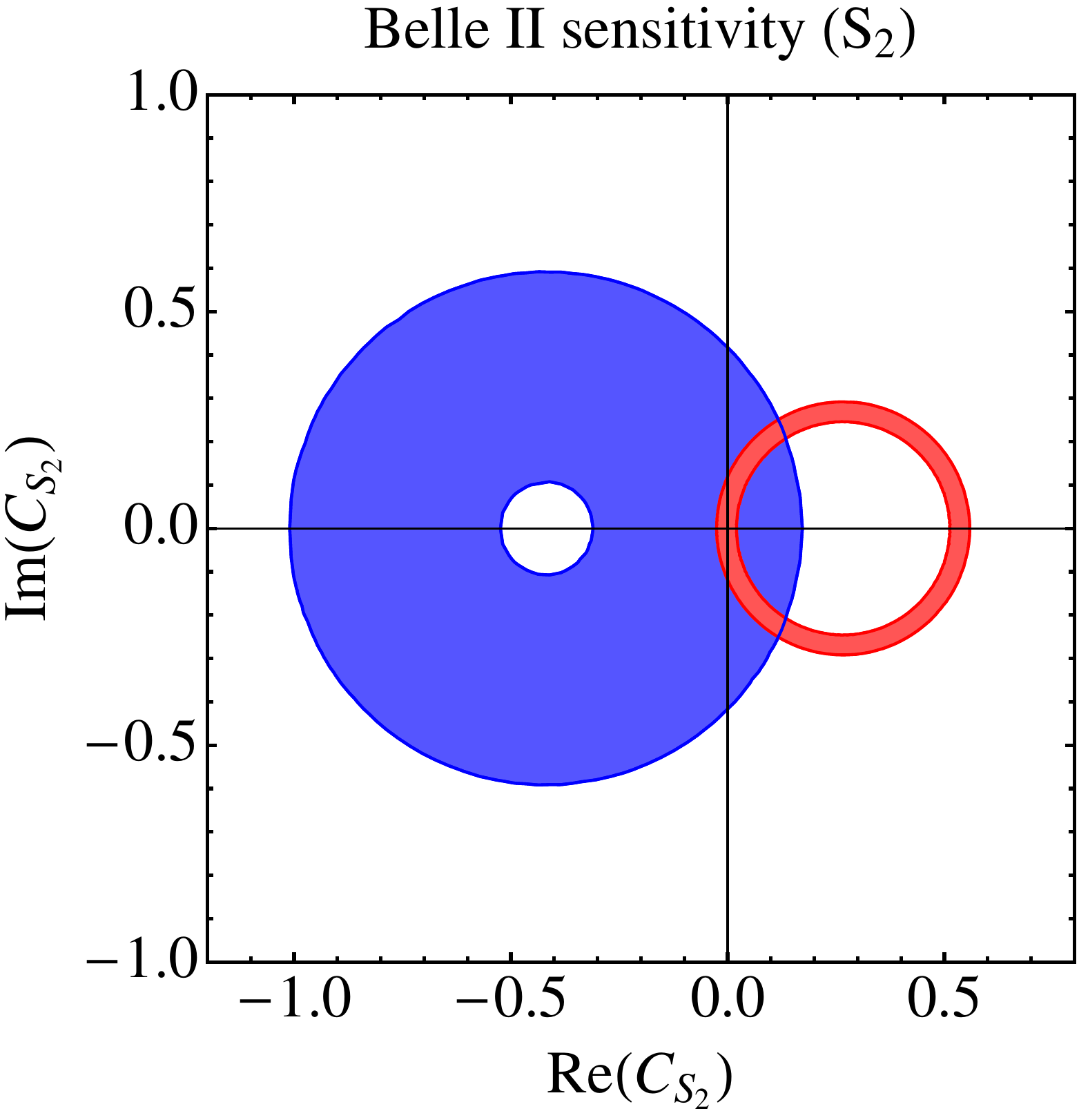} 
\caption{
Sensitivity to the new physics scenarios in terms of the 95\% CL allowed range of $C_X$ expected at the SuperKEKB/Belle~II with $50 \,\text{ab}^{-1}$ of accumulated data. 
The ``future'' experimental data are given as explained in the main text. 
A new physics contribution for the outside regions of the blue and red colors can be probed by $R_\pi$ and $R_\text{ps}$, respectively, at $50 \,\text{ab}^{-1}$ of Belle~II. 
} 
\label{Fig:Belle2Sensitivity}
\end{center}
\end{figure}
The 95\% CL expected constraints on the Wilson coefficients with these ``future'' experimental data are shown in Fig.~\ref{Fig:Belle2Sensitivity} for the $V_1$, $V_2$, $S_1$, and $S_2$ scenarios. 
A new physics contribution beyond the blue and red regions can be probed by measuring $R_\pi$ and $R_\text{ps}$, respectively at Belle~II. 
Each allowed region is annulus-like in the complex plane of $C_X$. 
For the $V_1$ scenario, the new physics sensitivities of $R_\pi$ and $R_\text{ps}$ are almost degenerate and the  region around $C_{V_1}\sim -2$ of large negative interference with the SM contribution is allowed by both of them. 
On the other hand, in the $V_2$ scenario, such regions of $C_{V_2}\sim -2$ for $R_\pi$ and $C_{V_2}\sim +2$ for $R_\text{ps}$ are incompatible with each other,  
as is already seen in the current constraint shown in Fig.~\ref{Fig:Bound}. 
For the scalar scenarios, the regions of $C_{S_1} =C_{S_2} \sim -0.8$ in $R_\pi$ and $C_{S_1} =-C_{S_2} \sim -0.5$ in $R_\text{ps}$ are of large negative interference. 
As is seen in the figures, we can test such a region for the $S_2$ scenario by combining $R_\pi$ and $R_\text{ps}$ while the sensitivity is relatively weak in the $S_1$ scenario. 
Therefore, the constraints from $R_\pi$ and $R_\text{ps}$ are complimentary and measuring both of them at SuperKEKB/Belle~II is meaningful to reduce allowed parameter regions, 
in particular for the $V_2$ and $S_2$ scenarios.

As for the tensor scenario, we also show the expected allowed region for $C_T$ in Fig.~\ref{Fig:BoundT}. 
The blue region with the dashed boundary indicates the one that can be tested with $50 \,\text{ab}^{-1}$, 
and the darker blue region with the dotted curve corresponds to the result for the case that the theoretical uncertainty in \eq{Eq:ThErrorT} is reduced by a factor of 2. 
As is explained in Sec.~\ref{SubSec:NPS}, the tensor scenario suffers from the larger theoretical uncertainty in $R_\pi$ 
so that we can see the significant effect of the reduction of the theoretical uncertainty.  
We also find that the present theoretical uncertainties for the vector and scalar scenarios are sufficiently smaller than the future (expected) experimental uncertainties\footnote
{The reduction of the theoretical error by factor 2, for example, gives only $0.1\%$ and $1\%$ differences in the expected allowed regions for the vector and scalar scenarios, respectively. 
}. 
We note that another observable such as $\mathcal{B}(B\to\rho\tau\bar\nu)$ is necessary to exclude the region of large negative interference of $C_T \sim -0.7$.

In Table~\ref{Tab:CXfromPureL}, we present the combined limits of the allowed ranges for $C_X$ (taken real) in order to quantify the expected sensitivities at SuperKEKB/Belle~II. 
It turns out that, focusing on the vicinity of the origin, the region of $|C_{X}| \gtrsim 0.03$ can be probed in the scalar scenarios. 
As for the vector and tensor scenarios, the Belle~II sensitivity is $|C_{X}| \sim 0.1$.

\begin{table}[t]
 \begin{center}
 \scalebox{0.8}{
 \begin{tabular}{cccc}
 \hline\hline
 NP scenario			&	 ~~$R_\pi^\text{Belle II}= 0.641\pm 0.071$ and $R_\text{ps}^\text{Belle II} =0.574 \pm 0.020$~~		&	~~$R_\text{pl}^\text{Belle II}=222 \pm 47$~~		  \\ 
 \hline
 $C_{V_1}$			&	$[-0.08, 0.09]; [-2.09, -1.92]$        						&	$[-0.23, 0.19]; [-2.19, -1.77]$		\\
 $C_{V_2}$			&	$[-0.09, 0.08]~~~~~~~~~~~~~~~~~~~~$       				 	&	$[-0.19, 0.23];~~~~~[1.77, 2.19]$		\\
 $C_{S_1}$			&	$[-0.03, 0.03]; [-0.55, -0.52]$       						&	$[-0.06, 0.05]; [-0.58, -0.47]$		\\
 $C_{S_2}$			&	$[-0.03, 0.03]~~~~~~~~~~~~~~~~~~~~$        					&	$[-0.05, 0.06];~~~~~[0.47,0.58]$		\\
 $C_{T}$				&	$[-0.13, 0.10]; [-1.23,-0.56]$			&	-        \\
 \hline\hline
 \end{tabular}
 }
 \caption{
Sensitivity to the new physics scenarios in terms of the 95\% CL allowed range of $C_X$ expected at the SuperKEKB/Belle~II with $50 \,\text{ab}^{-1}$ of accumulated data. The ``future'' experimental data are given as explained in the main text. The coefficient $C_X$ is assumed to be real.  }
 \label{Tab:CXfromPureL}
 \end{center}
\end{table}

The muonic mode $B \to \mu \bar\nu_\mu$ may also play an important role at SuperKEKB/Belle~II. 
At present, this process has not yet been observed and the current upper limit on the branching ratio is reported as 
$\mathcal B (B \to \mu \bar\nu_\mu)^\text{exp.} < 1 \times 10^{-6}$ at 90\% CL~\cite{Agashe:2014kda,Satoyama:2006xn,Aubert:2009ar}. 
This result may be compared with the SM prediction $\mathcal B (B \to \mu \bar\nu_\mu)^\text{SM} = (0.41 \pm 0.05) \times 10^{-6}$ 
and thus, we expect that $B \to \mu \bar\nu_\mu$ will be observed with a meaningful statistical significance at SuperKEKB/Belle~II. 
Accordingly, we introduce the pure-leptonic ratio
\begin{align}
 R_\text{pl} = \frac{ \mathcal B(B \to \tau \bar\nu_\tau) }{ \mathcal B (B \to \mu \bar\nu_\mu) }\,, 
\end{align}
as we defined $R_\pi$. 
In this paper, we assume contributions other than the SM do not exist in $B \to \mu \bar\nu_\mu$ as well as $B \to \pi \ell \bar\nu$. 
From the theory side, $R_\text{pl}$ is precisely evaluated as 
\begin{align}
 R_\text{pl} = \frac{m_\tau^2}{m_\mu^2} \frac{(1-m_\tau^2/m_B^2)^2}{(1-m_\mu^2/m_B^2)^2} | 1 +r_\text{NP} |^2 \simeq 222\, | 1 +r_\text{NP} |^2 \,. 
\end{align}
The dominant source of uncertainty $f_B |V_{ub}|$ in the leptonic decay rates cancels out and hence it is free from the $|V_{ub}|$ determinations, in which some discrepancies might still remain in the Belle~II era.

Following Ref.~\cite{Satoyama:2006xn}, the $1\sigma$ range of the error in $\mathcal B (B \to \mu \bar\nu_\mu)^\text{exp.}$ is obtained as $\pm 0.6 \times 10^{-6}$ at present. 
This is expected to be reduced as $\pm 0.08 \times 10^{-6}$ with $50 \,\text{ab}^{-1}$ at SuperKEKB/Belle~II. 
Applying the same procedure with $R_\pi$, namely with the expected ``future'' data being given as $R_\text{pl}^\text{Belle II}=222 \pm 47$, 
we have evaluated the future sensitivity of the ratio $R_\text{pl}$ to the new physics scenarios as shown in Table~\ref{Tab:CXfromPureL}. 
One finds that the sensitivity of $R_\text{pl}$ is rather ($\sim$ factor 2) weaker than that of $R_\text{ps}$. 
Although $R_\text{ps}$ has better performance, the ratio $R_\text{pl}$ is still a good observable 
in the sense that it has the very accurate theoretical prediction and could be used as a consistency check.

\section{Summary}
\label{Sec:Summary}

We have studied possible new physics in the semi- and pure- tauonic 
$B$ decays, $B \to \pi \tau\bar\nu_\tau$ and $B \to \tau \bar\nu_\tau$,
using the model-independent effective Lagrangian including the vector
($V_{1,2}$), scalar ($S_{1,2}$), and tensor ($T$) types of
interaction. The formulae of the differential branching fractions 
in the presence of new physics described by the effective Lagrangian 
are presented with a brief summary of the hadronic form factors
in the $B \to \pi$ transition.  

We have examined the ratio of the branching fraction of $B \to \pi
\tau\bar\nu_\tau$ to that of $B \to \pi \ell\bar\nu_\ell$, $R_\pi$ defined in
Eq.~\eqref{Eq:Rpi}, in order to reduce uncertainties in theoretical
calculations in analogy with $B \to D^{(*)}\tau\bar\nu_\tau$.  
Using the recent results of lattice QCD studies on the relevant form factors,
we have evaluated the effects of new physics in $R_\pi$ 
along with its theoretical uncertainty. 
The theoretical uncertainties in the $V_{1,2}$ scenarios are 
negligible compared to the present experimental error, and those in
the $S_{1,2}$ scenarios are sizable, but sufficiently small.
In contrast, the new physics contribution in the $T$ scenario
is rather uncertain as shown in Fig.~\ref{Fig:Effect}.

We have obtained the present constraints on the Wilson coefficients
that describe possible new physics contributions,
$C_X$ ($X=V_{1,2},S_{1,2},T$), comparing 
the theoretical predictions (with uncertainties mentioned above)
of $R_\pi$ and $R_\text{ps}$
with the experimental data. As shown in Fig.~\ref{Fig:Bound},
some of regions of $|C_X| \gtrsim O(1)$ are disfavored by
the current data. The sensitivity of $R_\pi$ in the $V_1$ scenario
is less than that of $R_\text{ps}$, but
their difference is not so significant. In the $V_2$ scenario, these
two observables probe different regions of $C_{V_2}$ and are
complementary. As for the $S_{1,2}$ scenarios, 
$R_\text{ps}$ is more sensitive
owing to the chiral enhancement. Since the tensor operator does not
contribute to $B \to \tau \bar\nu_\tau$, the $T$ scenario is 
constrained solely by $R_\pi$.
 
Furthermore, we have discussed the future prospect at the 
SuperKEKB/Belle~II experiment and shown its sensitivity to new physics 
in terms of expected constraints on $C_X$. 
Assuming that both the statistical and systematic uncertainties in the experiment are 
reduced as the integrated luminosity is increased to $50\ \text{ab}^{-1}$ 
and the central values are given by the SM, we have estimated
the expected allowed ranges of $C_X$ from $R_\pi$ and $R_\text{ps}$. 

It turns out that the allowed regions of $C_X$ are significantly reduced in all the scenarios and 
the region of large negative interference with the SM can be excluded by combining $R_\pi$ and $R_\text{ps}$ in the $V_2$ and $S_2$ scenarios as shown in Figs.~\ref{Fig:BoundT} and \ref{Fig:Belle2Sensitivity}.  
The SuperKEKB/Belle~II experiment can probe the new physics contribution of $|C_{X}|$ as small as $0.03$ in the scalar scenarios and $\sim 0.1$ in the vector and tensor scenarios as seen in Table~\ref{Tab:CXfromPureL}.

Further improvement of sensitivity may be achieved if $R_\pi$ and 
$R_\text{pl}$ are measured by a similar method adopted to 
measure $R_{D^{(*)}}$, namely not separate measurements of the numerator
and denominator but direct measurements of the ratios. 
It is also desired to improve the precision of the tensor form factor
as well as to evaluate the scalar form factor by lattice simulation.
The latter is useful to eliminate the potential uncertainty in 
the bottom quark mass arising from the equation of motion. 
Supplemental observables such as $\mathcal{B}(B\to\rho\tau\bar\nu)$ 
and the $q^2$ distribution of $B\to\pi\tau\bar\nu$ are also helpful
to further squeeze $C_X$ as well as to probe or exclude the region of negative interference in the $S_1$ and $T$ scenarios.

\section*{Acknowledgment}
We are grateful to Florian Bernlochner for his useful comments on the $B\to\pi\tau\bar\nu_\tau$ data and 
Hidenori Fukaya for his suggestion on the lattice QCD result. We also thank
Tetsuya Enomoto for discussions in the early stage of 
this study. This work is supported in part by JSPS KAKENHI 
Grant Numbers JP25400257 and JP16H03993 (MT), and IBS-R018-D1 (RW).

\bibliographystyle{ptephy}
\bibliography{reference_ETW}

\end{document}